**Ultra-thin SWNTs Films with Tunable, Anisotropic Transport Properties**

By *Bo Li, Hyun Young Jung, Hailong Wang, Young Lae Kim, Taehoon Kim, Myung Gwan Hahm, Ahmed Busnaina, Moneesh Upmanyu\*, and Yung Joon Jung \**

[*]     Prof. Y. J. Jung. Corresponding-Author, B. Li, Dr. H. Y. Jung, Dr. T. Kim, Dr. M. G. Hahm, Prof. A. Busnaina
Department of Mechanical and Industrial Engineering,
Northeastern University, Boston, MA 02115(USA)
E-mail: jungy@coe.neu.edu
        Prof. M. Upmanyu, Corresponding-Author, Dr. H. Wang
Group for Simulation and Theory of Atomic-scale Material Phenomena (stAMP),
Northeastern University, Boston, MA 02115 (USA)
E-mail: m.upmanyu@neu.edu
        Y. L. Kim
Department of Electrical and Computer Engineering,
Northeastern University, Boston, MA 02115 (USA)

Directional transport properties at the nanoscale remain a challenge primarily due to issues associated with control over the underlying anisotropy and scalability to macroscopic scales. In this letter, we develop a facile approach based on template-guided fluidic assembly of high mobility building blocks - single walled carbon nanotubes (SWNTs) - to fabricate ultra-thin and anisotropic SWNT films. A major advancement is the complete control over the anisotropy in the assembled nanostructure, realized by three-dimensional engineering of dip-coated SWNT thin films into alternating hydrophilic and hydrophobic micro-line patterns with prescribed intra/inter-line widths and line thicknesses. Variations in the contact line profile results in an evaporation-controlled assembly mechanism that leads to the formation of an alternating, and more importantly, contiguous SWNT network. Evidently, the nanoscopic thickness modulations are direct reflections of the substrate geometry and chemistry. The nanostructured film exhibits significant anisotropy in their electrical and thermal transport properties as well as optical transparency, as revealed by characterization studies. The direct interplay between the anisotropy and the 3D micro-line patterns of the substrate combined with the wafer-level scalability of the fluidic assembly allows us to tune the transport properties for a host of nanoelectronic applications.

# 1. Introduction

Nanoelectronic devices that rely on the superior electronic transport in individual single-walled carbon nanotubes (SWNTs) are typically beleaguered by reliability and scalability issues, the primary reason for the ongoing paradigm shift towards SWNTs thin films as active elements.[1] Recent success in wafer-scale synthesis of these thin films, either by SWNT growth or by assembly, [1-6] has emerged as the driving force for several applications including next-generation flexible and transparent thin-film electrodes, high mobility transistors, and a suite of robust sensors. Expectedly, engineering the carbon nanotube-scale nanostructure as well as the overall film architecture is the key in tailoring electronic transport through the films. Unfortunately, it remains a limitation as the existing fabrication techniques typically yield one- or two-dimensional (1D/2D) isotropic SWNT network film structures.[3, 7, 8] Compounding matters is the fact that the constituent SWNTs are usually a mixture of semiconducting and metallic nanotubes that cannot be easily separated.[9, 10] Interestingly, though, the interplay between the intrinsic electrical heterogeneity and the nanoscale structure/topology of the assembled SWNT networks can result in drastically different transport characteristics[11-13] and therefore is a promising route for engineering highly functional and integrated film systems with prescribed electrical, thermal, and optical properties.

Tailoring the nanotube-scale network in turn requires complete control over film density, morphology and architecture. In this letter we present a facile route for the synthesis of ultra-thin SWNT film architectures which permit engineering of the (in-plane) directionality in the electrical and thermal transport, i.e. the anisotropy. The enabling component unique to our approach is the fluidic assembly of dispersed SWNTs onto suitably micro-patterned and chemically heterogeneous three-dimensional (3D) substrates. The resultant SWNT films have several features that allow control over the extent of the anisotropy. First, it is immediately clear that the assembled SWNTs inherit the topography of 3D patterned SiO$_2$ and photoresist (PR) micro-line structures, forming quasi-wavy, ultra-thin, and transparent films. Second, the chemical heterogeneity of the substrate due to the alternating hydrophobic (PR) and hydrophilic (SiO$_2$) micro-patterns allow us to control the local thickness of SWNT films, thereby allowing control over the electrical and thermal properties. Lastly, these 3D SWNT films can be easily transferred onto polymer matrices, a critical step in synthesis of flexible and transparent SWNT network-polymer composite films for flexible, transparent and anisotropic components in electrical devices.

## 2. Results and Discussion

### 2.1. Morphology of 3D SWNT Films

Figure 1 shows representative scanning electron microscopy (SEM) and optical images as well as schematics of our assembled SWNT films. To this end, the underlying SiO$_2$ or quartz substrates are first subjected to plasma treatment to improve their affinity to SWNT-deionized (DI) water solution.[14-16] Then, chemical heterogeneity is introduced through the fabrication of hydrophobic PR micro-line structures (6 μm in width with a spacing of 9 μm) with controlled PR thicknesses ($H_{PR}$=323, 360, and 480 nm) using photolithography processes, resulting in a 3D hydrophobic-hydrophilic surface architecture. Finally, the fluidic assembly of SWNTs is realized by vertically dip-coating the substrate into 0.23 wt% SWNT-DI water solution and gradually lifting at a controlled pulling velocity of $V$=0.1 mm/min. SEM images (Fig. 1a and 1b) clearly show that the assembled SWNT network forms a continuous thin film which inherits the 3D morphology of hydrophobic and hydrophilic micro-lines on the substrate, i.e. the film morphology is quasi-wavy as illustrated schematically in Fig. 1c. This assembled SWNTs film is transparent as well, evident from the optical microscope image shown in Fig. 1d. We were also able to assemble these ultra-thin 3D SWNT films on PR/quartz substrates. Figure 1e shows the optical image of the sample after removing the PR, further illustrating the transparent nature of our assembled SWNT films.

A notable feature, revealed by atomic force microscopy (AFM) characterization in Figure 2, is that the local thickness of these assembled SWNT films differs dramatically between the PR and $SiO_2$ micro-lines, resulting in alternating thin and thick SWNT film strips. The inset schematic in Fig. 2 shows the cross sectional view of the 3D SWNT films on the patterned substrate. A key aspect that allows us to engineer the SWNT film morphology is that the thickness of SWNT films on photoresist ($H_{SWNT/PR}$) is strongly influenced by $H_{PR}$. More specifically, lower $H_{PR}$ (323 nm) results in a higher $H_{SWNT/PR}$ (40 nm). Conversely for higher $H_{PR}$ (480 nm), the film thickness decreases to a few nanometers, $H_{SWNT/PR}$=5.5 nm, while the thickness of SWNTs film on $SiO_2$ ($H_{SWNT/SiO2}$) is relative inert to $H_{PR}$.

## 2.2. Characterization of Anisotropic Transport Properties

Controlled modulations in the local thickness of the assembled SWNTs films can be exploited to tailor the directionality of the electrical transport through SWNTs film. To this end, the 100×100 μm² square gold contact pads were patterned on the SWNTs film with 100μm spacing, as shown in Figure 3a (inset). The electrical resistances were measured in the vertical V-V' ($R_{vertical}$) and in the horizontal H-H' direction ($R_{horizontal}$) by two-point probe method. The dependence of the electrical resistance ($R_{vertical}$, red line and $R_{horizontal}$, green line) as well as the electrical anisotropy ($H_{SWNT/SiO2}/H_{SWNT/PR}$, blue line) on thickness ratio of SWNT film ($H_{SWNT/SiO2}/H_{SWNT/PR}$) are shown in Fig. 3. For all samples, $R_{horizontal}$ is higher than $R_{vertical}$, strongly indicative of the in-plane electrical anisotropy in these films. Furthermore, the electrical resistance in both directions increased with the increase in the ratio $H_{SWNT/SiO_2}/H_{SWNT/PR}$, or increase in the PR thickness. However, $R_{horizontal}$ increases more rapidly such that the electrical anisotropy, expressed as $R_{horizontal}/R_{vertical}$, also increases with $H_{SWNT/SiO_2}/H_{SWNT/PR}$. The results clearly demonstrate that the electrical anisotropy can be controlled by changing the thickness of photoresist $H_{PR}$. Simple percolation based simulations of randomly assembled SWNT networks corroborate this effect, as shown in Figure S1. The inverse number of possible conductive paths (1/paths), related to the resistance of the film, increase with the increase of inverse number of stacking layer (1/layers), or the thickness of film. Decreasing film thickness leads to a non-linear decrease of possible conductive path, and thus a non-linear increase of resistance. The effect is a reflection of the reduction in effective conduction pathway of metallic SWNTs in assembled SWNTs films.[13] In addition, owing to the electrical heterogeneity of the SWCNT mix, the metallic SWCNTs get effectively shielded such that the nature of electrical conduction becomes increasingly semiconducting at small thicknesses.[13] It follows then that in the 3D anisotropic system with variations in film thickness, the electrons pass the thin and thick SWNT strips in parallel along the vertical direction and the thicker films dominate the electron transportation. However, in the horizontal direction, the electrons pass alternating thin and thick strips in series such that thinner SWNT strips determine the overall electrical transport property of these SWNTs films. In each case, the vertical direction (V-V') has enhanced conduction pathways relative to the horizontal direction.

In the vertical direction, the electrical resistance is mainly determined by the thick SWNT strips on $SiO_2$ alongwith the thin SWNT strips on PR that provide extra conductive paths in parallel. Thus, as the height ratio increases, the number of conducting paths reduce as the the metallic SWCTs in the heterogeneous SWNT mixture are shielded such that the resistance along the PR trenches is increasingly controlled by semiconducting SWNTs. The combination of geometry and nature of electrical conduction drives the resistance of the PR film ; as mentioned earlier, the increase is highly non-linear (see Supplementary Documents and Ref[13]). Consequently, the resistance of the SWCT film on the PR becomes significant and drives the overall resistance. This is evident in Fig. 2 wherein the decrease in $H_{SWNTs/PR}$ due to increasing PR height results in higher vertical resistance. Additional effects due to variations in the $SiO_2$ trench geometry and film height as well as non-trivial structures at the PR-$SiO_2$ interface cannot be ruled out. Their characterization, however, is beyond the scope of this study.

We also investigated thermal transport properties of the assembled SWNTs film having the thickness ratio ($H_{SWNT/SiO2}/H_{SWNT/PR}$) of 1.44. The measurement of thermal conductivity of the anisotropic SWNTs film was performed by utilizing a self-heating $3\omega$ technique (see Figs. 3b and 3c). The $3\omega$ signal correlates with thermal conductivity through Eq.1,

$$V_{3\omega, rms} = \frac{4 I^3 R R' L}{\pi^4 k S}, \quad (1)$$

where *L, R,* and *S* are the distance between contacts, electrical resistance, and cross sectional area of the sample, respectively. *R′=(δR/δT)* is the temperature gradient of the resistance at the chosen temperature and *k* is the thermal conductivity.[17, 18] The $3\omega$ method was utilized by the four-point-probe third harmonic characterization to eliminate the contact resistance and to avoid related spurious signals. Specifically, the resistance of the assembled SWNTs film and its temperature dependency in the vertical and horizontal directions was measured in the temperature range of 21 ~ 25 °C which lies close to the measurement temperature of the $3\omega$ signal. The measured resistance and the temperature coefficient are 422 Ω and 31 Ω/°C along the vertical direction (V-V'), and 469 Ω and 25 Ω/°C along the horizontal direction (H-H') (Fig. 3b). Again, the lower temperature coefficient in horizontal direction results from the reduction in effective conduction pathway of metallic SWNTs. The third-harmonic voltage measured at the frequency of 1000 Hz is shown in Fig. 3c. The thermal conductivities calculated with Eq.1 were 22 W/mK and 4 W/mK along the vertical and horizontal directions, respectively. We assume that the decrease in the effective thermal conductivity across the horizontal direction is due to the reduced conduction pathways relative to the vertical direction, and additional phonon scattering at the interfaces between the alternating thin and thick film strips.[19]

### 2.3. Transfer of the 3D SWNT film

We also demonstrate the ability to transfer assembled SWNT films on flexible polymeric substrates, a crucial capability for the synthesis of scalable yet tailored integrated functional flexible systems. Figure 4a shows the schematic of developed transfer process: first, a layer of poly(methyl methacrylate) (PMMA) is spin-coated on the SWNT films; second, the underlying $SiO_2$ layer is etched by HF solution; finally, the PR/SWNT/PMMA film is peeled off from the original substrate and can be readily transferred onto any other target substrate. Due to the unique 3D morphology, our assembled films can be transferred together with arrays of PR micro-lines inserted into the trenches of 3D SWNT films. Figure 4b shows the SEM image of the transferred SWNT film, in which the black color regions are exposed to SWNT film while the bright color regions are PR micro-lines. A well defined boundary between $SiO_2$ and PR micro-lines can be seen in Fig. 4c. Finally as confirmation, Fig. 4d shows the optical image of the transferred PR/SWNT/ PMMA hybrid film.

## 2.4. Formation Mechanism of 3D SWNT films

To elucidate the formation mechanism of anisotropic 3D SWNT films, we have first quantified the chemical heterogeneity of the substrate with respect to the SWCNT-DI solution via micron-scale static contact angle measurements. Our results reveal (Fig. 6a) that the PR is significantly hydrophobic ($\theta$=60° for PR-solution) compared to the plasma treated $SiO_2$ ($\theta$<5°). In our experiments, the SWNTs deposit along the receding liquid-air contact line. Then, the interplay between the chemistry and morphology of the substrate and the microfluidics of the receding contact line becomes important,[20, 21] evident from the snapshot of the assembly during the dip-coat shown in Fig. 5b. The SEM image above the reservoir is along the film normal while that below is a schematic illustration. The wavy contact line, consisting of liquid bridges across the PR lines that are curved along the dip-coat direction, is a result of a solution that wets into the PR, strongly suggestive of "cross-talk" between the micro-lines, i.e. modification of the contact line dynamics due to chemical heterogeneity of the abutting lines. While its origin follows from the simple fact that the width of the hydrophobic PR pattern lines (~6 μm) is much smaller than the capillary length of the SWNT solution (see Supplementary Documents for a discussion), the effect on the final film thickness on both PR and $SiO_2$ micro-lines requires an understanding of the microfluidics during the dip-coat. To illustrate the salient effects, consider a cartesian coordinate system with its *xy* plane along the substrate (Fig. 5c): *x* denotes the direction of plane withdrawal (streamwise direction) and *z* is the direction normal to the plane. The large curvature transverse to the pattern lines and the change in contact angle (~5° to 60°) at the chemically heterogeneous interface between the patterns modify the excess pressure at the liquid-air interface and therefore the shape of the meniscus *h*(*x, y*) associated with the entrained liquid. The excess pressure is central to the extent of the cross-talk between the patterns and for small slopes it can be approximated as $\delta p = \gamma(h_{xx} + h_{yy})$. Near the reservoir, the much larger $h_{xx}$ is controlled by the width of the hydrophobic pattern lines $W$ and scales as $h_{xx} \sim \delta/W^2$ at the center of PR pattern, where $\delta$ is a length scale along the streamwise direction. Evidently, large capillary pressures are required to balance the excess pressure and drive the enhanced liquid pick-up into the PR micro-lines forming a wetting layer well above the horizontal of the reservoir (Fig. 5b and 5c).

The above analysis elucidates the basis for the formation of continuous SWNT thin films, yet the fluidics that controls the local thickness of the film is slightly different in that, i) for the set of deposition variables employed in this study the film thickness is evaporation-controlled, and ii) the variation in the PR step thickness (Fig. 5c) introduces an additional transverse curvature $h_{zz}$ due to the out-of plane variation in the thickness profile, $h \equiv h(y, z)$. In the absence of cross-talk, the film thickness following evaporation is $H \cong \Phi \theta l_d$, where $\Phi$ is strength of the solution and $l_d$ is the drying length, the length along the streamwise direction at which the SWNT flux due to motion of the contact line is exactly balanced by the evaporation flux. Following Berteloot,[22] a one-dimensional flux balance yields $l_d = (J_0/V\theta)^2$, where $J_0$ is evaporation constant for the solvent (approximately that of water) which depends on ambient conditions such as vapor saturation concentration, diffusion constant, mass density, etc. Under standard conditions, $J_0 = 10^{-9}$ m$^{3/2}$s$^{-1}$ for a millimeter sized water droplet. For the dip-coating parameters in our experiments, this yields a drying length of $l_d \sim 4.7$ μm and $l_d \sim 0.3$ μm for the SiO$_2$ and PR surfaces, respectively; the corresponding film thicknesses on SiO$_2$ and PR are $\sim 9.5$ nm and $\sim 0.8$ nm (see Supplementary Documents). The estimates for film thicknesses represent lower bounds as they correspond to infinite, homogeneous surfaces. The larger thicknesses reported in Fig. 5c, especially on PR lines, are in part an indicator of the cross-talk which leads to continuous thin film formation. The traverse flow decreases the effective velocity of the receding contact line on the PR lines, thereby increasing the drying length and the evaporation-controlled film thickness ($\sim 1/V^2$). The change in the thickness of the SiO$_2$ lines is minimal as the contact line profile lags away from the reservoir such that the transverse flow away from the line occurs at a region where viscous forces become important. The observed trends in the film thickness are due to an out-of-plane curvature that increases with PR step height and works against the in-plane transverse curvature, therefore opposing the liquid pick-up from the SiO$_2$ lines onto the PR surface (see Fig. 5c). In the evaporation-controlled limit, this leads to a reduction in the mean local velocity of the entrained liquid relative to the PR surface. As discussed before, we expect minimal changes in the thickness deposited on SiO$_2$ lines, consistent with our data. In summary, the thickness of the PR region, $H_{PR}$, emerges as a robust knob that offers complete control over the SWNTs film thickness, and the form and extent of the heterogeneity in film morphology.

## 3. Conclusions

We have successfully demonstrated a scalable yet facile route for fabricating highly engineered ultra-thin anisotropic SWNT films formed on 3D micro-patterned substrates using a highly controlled template guided fluidic assembly technique. Our assembled ultra thin SWNT film structures show continuous but alternating arrangement of SWNT line patterns with differing thicknesses. The thickness ratio can be controlled to at least an order of magnitude using our assembly technique. A simple scaling analysis explains the fundamental assembly mechanism that allows us to control the nanostructural heterogeneity. The electrical and thermal anisotropy is a direct result of the anisotropy in the nanostructure, and therefore can be tuned by controlling the 3D structure of the substrate. Moreover, by using a polymer casting transfer method, the films can be transferred onto the PMMA substrate to form heterogeneous composite films with multiple polymeric materials. The anisotropic and composite SWNT films with tunable electrical and thermal transport properties are expected to have broad implications for the development of next-generation flexible and transparent thin-film electrode, transistors and sensors, with minimized device-to-device variation and enhanced stability.

## 4. Experimental

*Measuring the thickness of 3D SWNTs films.* Atomic force microscopy (PSIA XE-150, Park Systems Inc., USA) was employed to measure the thickness of 3D SWNTs films. First, the profile of photoresist (PR) patterns on silicon dioxide ($SiO_2$) substrate was determined and the height of photoresist ($H_{PR}$) was obtained prior to the fabrication of SWNTs films. After depositing the 3D SWNTs film, the part of SWNTs film was carefully removed by a tungsten probe to expose $SiO_2$ substrate without any damage on the substrate. The exposed $SiO_2$ surface was used as the reference plane. Then, a large area scanning (30 μm x 15 μm) was performed to determine the thickness of SWNT film on $SiO_2$ surface ($H_{SWNT/SiO2}$) as well as the height of the PR with coated SWNTs film ($H_{PR+\,SWNT/PR}$). Finally, the thickness of SWNT film on PR ($H_{SWNT/PR}$) was obtained by subtracting $H_{PR}$ from the $H_{PR+SWNT/PR}$.

*Thermal conductive measurement.* Heat transport equation can be expressed in terms of the third harmonic voltage signal induced by an AC current of the form $V_{3\omega} = I_o \sin\omega t$ passing through the sample at low frequencies. The AC current with frequency $\omega$ creates a temperature fluctuation at $2\omega$, which further causes a third harmonic voltage signal. The $3\omega$ signals can be used for measuring thermal conductivity of anisotropic SWNT film. A lock-in amplifier (Stanford Research System SR850) was used for obtaining $3\omega$ signals by amplifying the small voltage and removing the noise. An AC current source (Keithley 6221) was used to provide a stable current supply. All the measurements including resistance, temperature, and 3ω signals were done under high vacuum (P<10$^{-5}$ Torr) in a Janis Research ST-500 cryogenic probe station to reduce radial heat losses through gas convection. The temperature coefficient of the resistance is also measured in order to obtain the thermal conductivity based on Eq.1. The thermal conductivities calculated with Eq.1 were obtained to be 22 W m$^{-1}$ K$^{-1}$ and 4 W m$^{-1}$ K$^{-1}$ for vertical and horizontal direction sample, respectively.

**Acknowledgements**

We would like to thank financial support from NSF CMMI Nanomanufacturing Program (0927088), Center for High-Rate Nanomanufacturing in Northeastern University, and Fundamental R&D Program for Core Technology of materials by Ministry of Knowledge Economy, Republic of Korea. We are also grateful to Brewer Science Corp. for kindly donating the relevant SWNT solutions. HW and MU are grateful for support from Structural Metallics Program, ONR -N000141010866.

Received: ((will be filled in by the editorial staff))
Revised: ((will be filled in by the editorial staff))
Published online: ((will be filled in by the editorial staff))

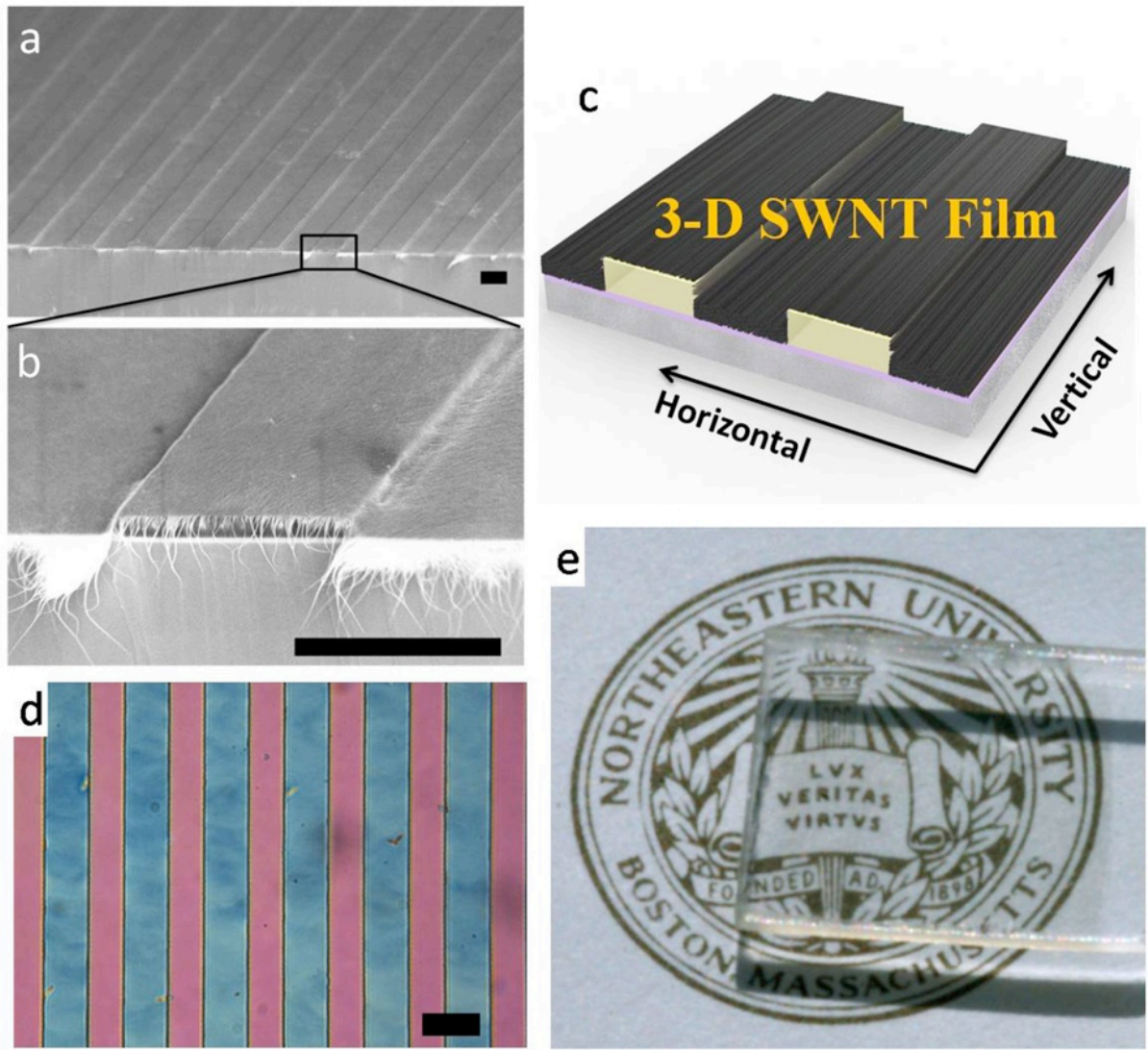

**Figure 1.** (a) A tilted SEM of assembled SWNTs film formed on photoresist (PR) micro-lines/ SiO$_2$ substrate. The scale bar is 5 μm and the thickness of PR is 360 nm. (b) An enlarged SEM image showing quasi-wavy 3D morphology of ultra-thin SWNTs film. The scale bar is 5 μm. (c) Schematic illustrating the anisotropic and nanostructured of SWNT film. (d) Optical microscopy image of the SWNT film formed on PR/SiO$_2$ substrate. Note that the background colors can be observed directly (pink and blue colors representing the PR and SiO$_2$ strips, respectively) due to the transparent nature of the developed SWNT film. The scale bar is 10 μm. (e) An optical picture of SWNT film assembled on the quartz substrate after removing the PR.

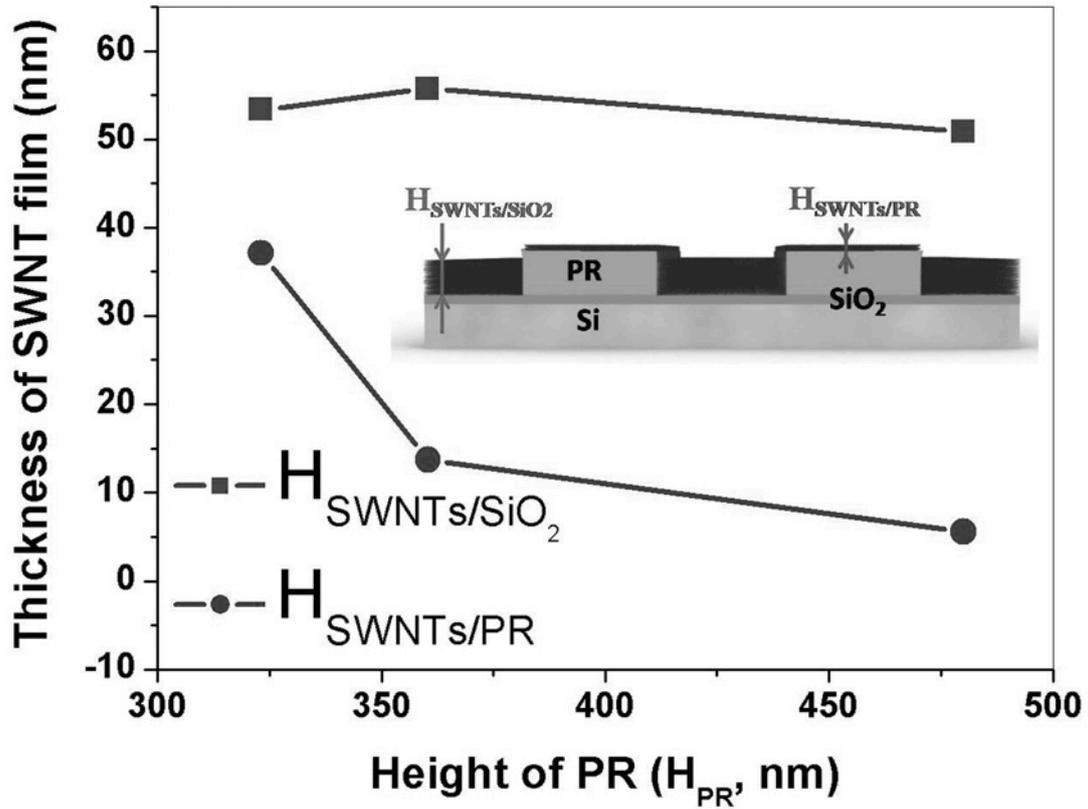

**Figure 2.** The dependence of the film heights $H_{SWNT/SiO2}$ (square line), $H_{SWNT/PR}$ (dot line) on the PR height $H_{PR}$. $H_{SWNT/PR}$ decreases with increasing $H_{PR}$, while $H_{SWNT/SiO2}$ is constant. (inset) Schematic of the cross section of 3D SWNT film with continuous but alternating arrangement of thin and thick SWNT strips.

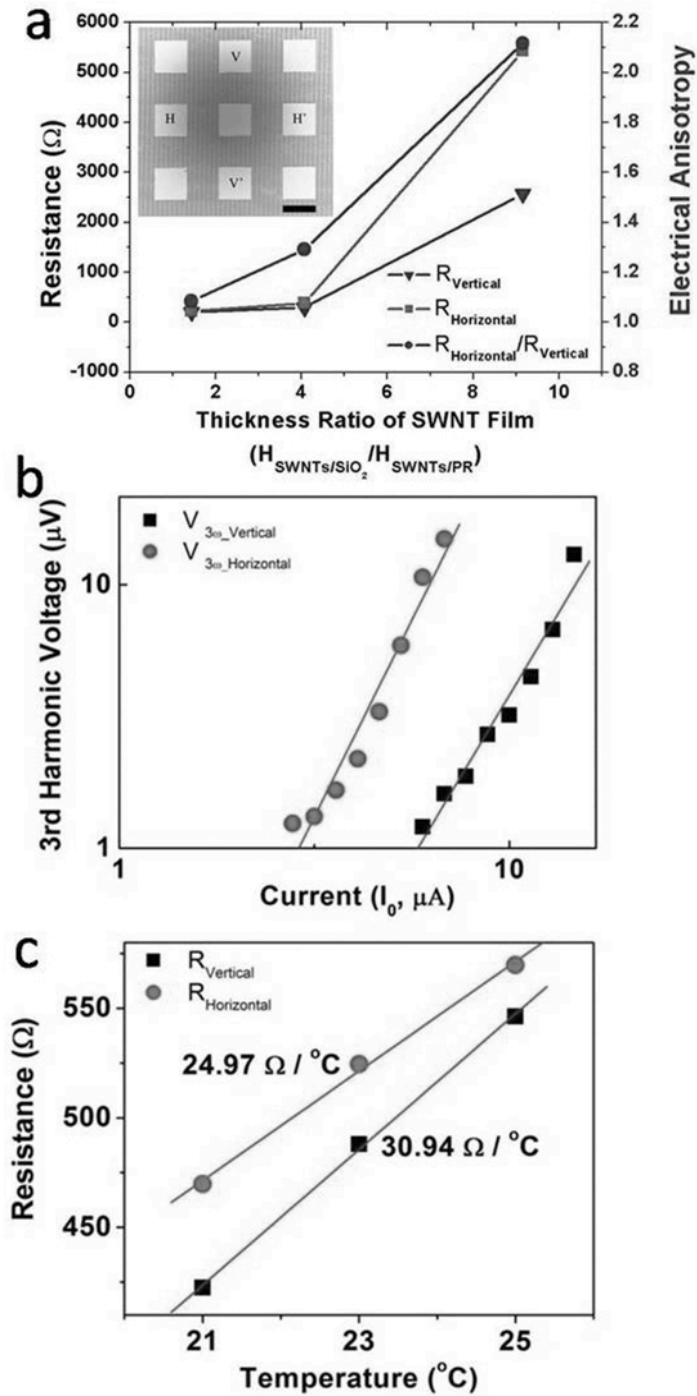

**Figure 3.** (a) The dependence of $R_{horizontal}$ (square line), $R_{vertical}$ (triangle line), and the electrical anisotropy ($R_{vertical}/R_{horizontal}$) on the thickness ratio of SWNTs film ($H_{SWNT/SiO_2}/H_{SWNT/PR}$). (inset) SEM image of a 3D SWNT film patterned with the 100×100 μm² square gold contact pads. (b) Temperature coefficient measurement in vertical and horizontal directions of the anisotropic SWNTs film at $T$=21~25 °C. The electrical resistance linearly

increases as the temperature rises. The calculated temperature coefficients for vertical and horizontal direction are 31 Ω/°C and 25 Ω/°C, respectively. (c) 3$\omega$ amplitude as a function of input current at a working frequency of 1000 Hz.

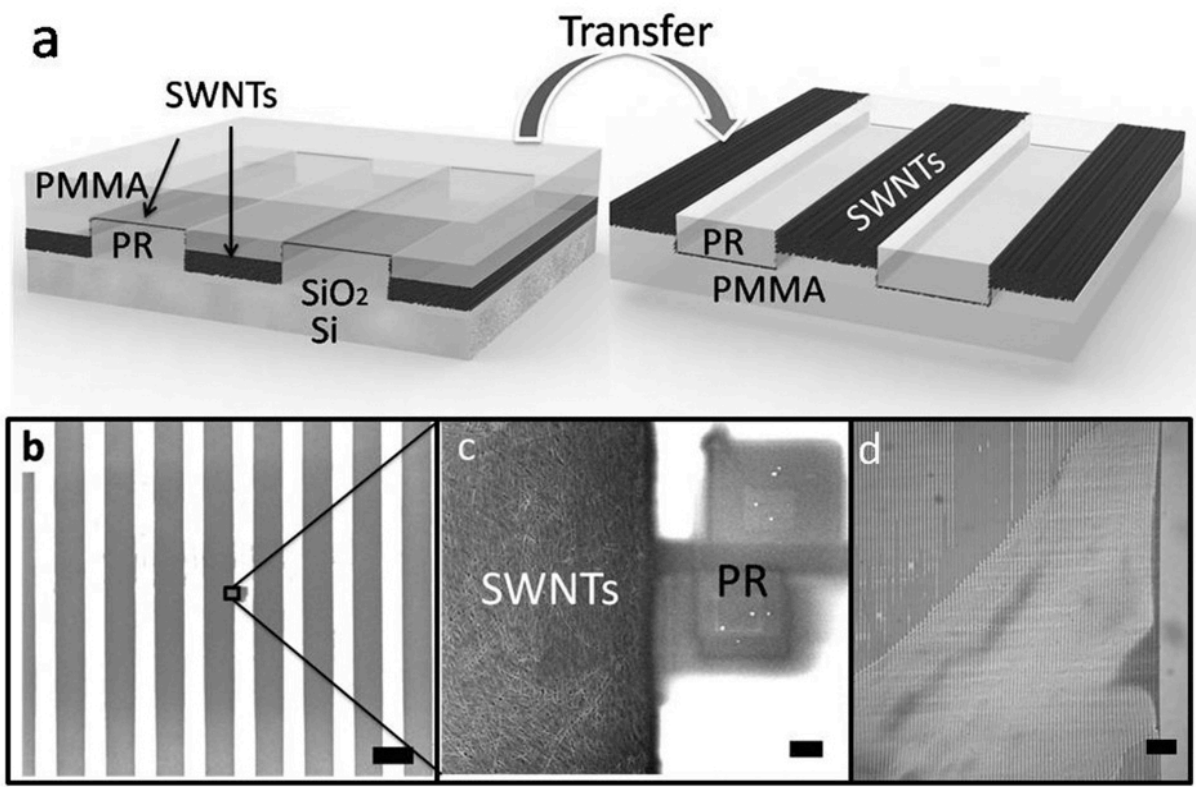

**Figure 4.** (a) Schematic illustration of the transfer of ultra-thin and anisotropic SWNT films to form a SWNT/PR/PMMA hybrid composite structure. (b) A SEM image of the transferred composite film. The scale bar is 10 μm. (c) Enlarged SEM image showing well defined boundary between SWNT film and PR. Note that PR lines are inserted on trenches of the film which is then placed on the thin PMMA layer. The scale bar is 200 nm. (d) The optical micrograph showing the transferred SWNTs/PR/PMMA hybrid composite film. The scale bar is 100 μm.

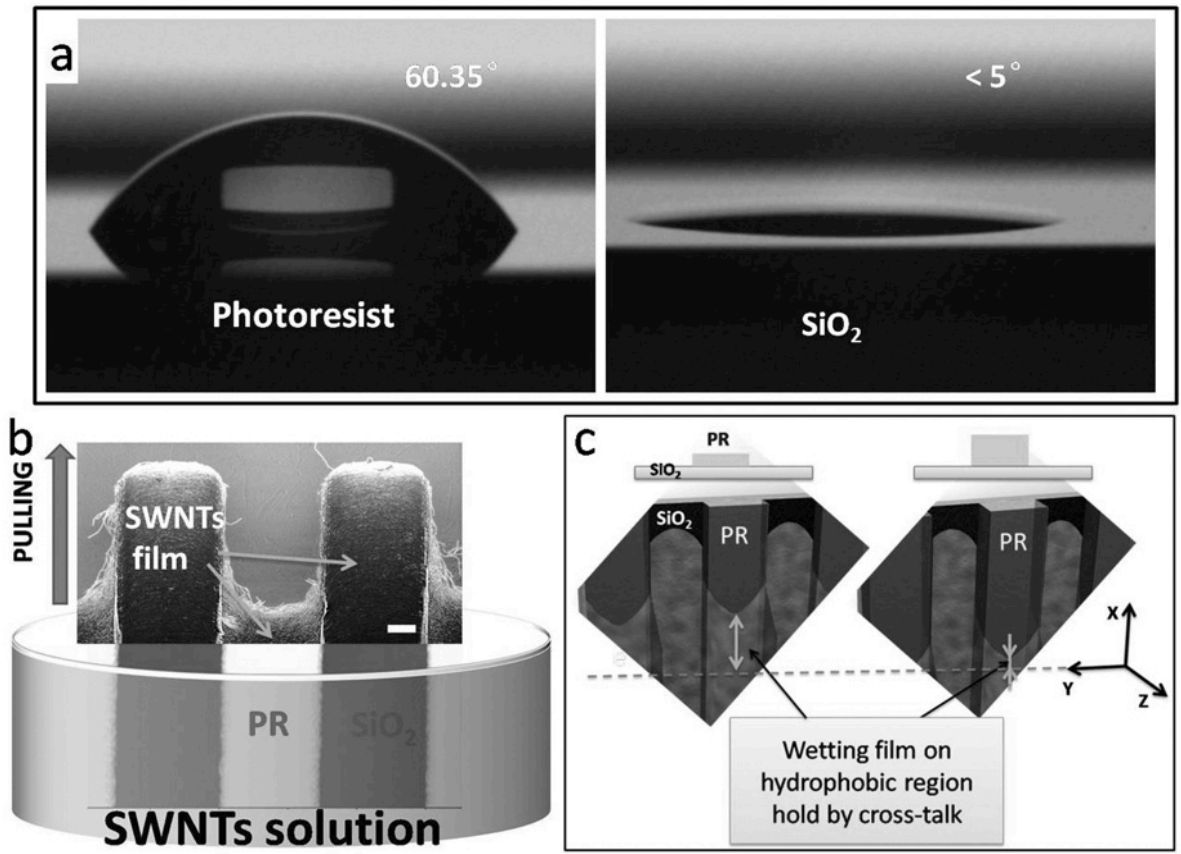

**Figure 5.** (a) Contact angle measurements of SWNTs/DI water solution on PR (left) and plasma treated SiO$_2$ surface (right). (b) Schematic showing the fluidic assembly of SWNTs through a dip coating process. SEM image above the horizontal of the SWNTs reservoir solution was taken from the top edge of the film and therefore indicates the initial point when the continuous SWNTs film was formed. The scale bar is 2 μm. (c) Schematic of the liquid-solid contact line formed on low H$_{PR}$ (left) and high H$_{PR}$ (right). A cartesian coordinate system is defined: *x* denotes the direction of plane withdrawal; *y* denotes the in-plane direction perpendicular to the *x* direction and *z* denotes the direction normal to the plane.

Submitted to **ADVANCED FUNCTIONAL MATERIALS**The table of contents entry

Anisotropic single walled carbon nanotube film was fabricated by template-guided fluidic assembly with complete control over the local thickness in the assembled nanostructure, realized by three-dimensional engineering of assembled substrate with alternating hydrophilic and hydrophobic micro-line. The nanostructured film exhibits significant in-plane anisotropy in the electrical and thermal transport properties as well as optically transparent nature.

Keywords

Single walled carbon nanotube, thin film, anisotropy, transfer.

B. Li, H. Y. Jung, H. Wang, Y. L. Kim, T. Kim, M. G. Hahm, A. Busnaina, M. Upmanyu*, and Y. J. Jung*

Title
Ultra-thin SWNTs Films with Tunable, Anisotropic Transport Properties

ToC figure

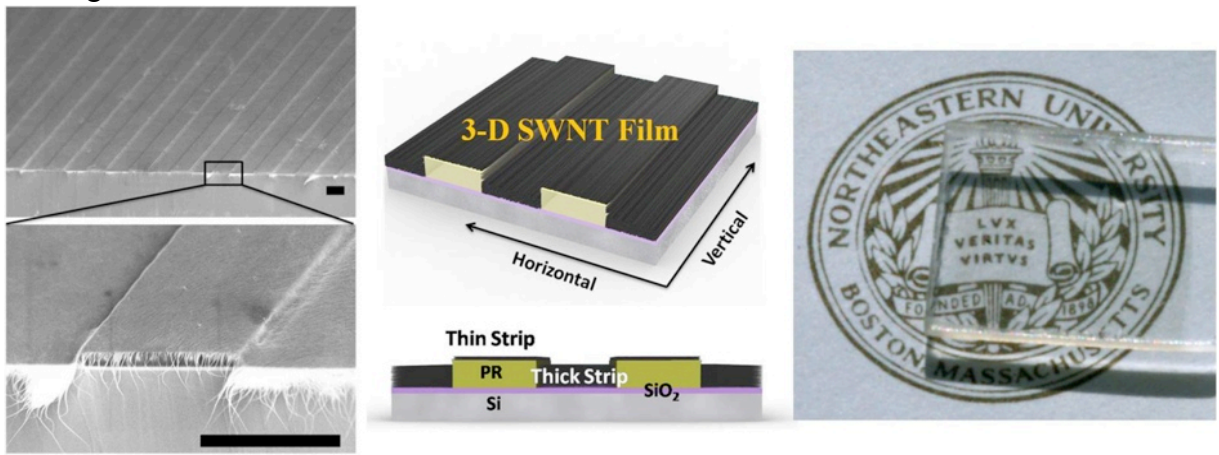